\documentstyle[12pt]{article}
\begin{document}
\date{}
\title{The Effective Action for Local Composite Operators
$\Phi^2(x)$ and $\Phi^4(x)$}
\author{Anna Okopi\'nska\\ 
Institute of Physics, Warsaw University, Bia\l ystok Branch,\\ 
Lipowa 41, 15-424 Bia\l ystok, Poland\\e-mail: rozynek@fuw.edu.pl}
\maketitle
\thispagestyle{empty}
\begin{abstract}
\noindent 
The generating functionals for the local composite operators,
$\Phi^2(x)$ and $\Phi^4(x)$, are used to study excitations in
the scalar quantum field theory with $\lambda \Phi^4$
interaction. The effective action for the composite operators is
obtained as a series in the Planck constant $\hbar$, and the
two- and four-particle propagators are derived. The numerical
results are studied in the space-time of one dimension, when the
theory is equivalent to the quantum mechanics of an anharmonic
oscillator. The effective potential and the poles of the
composite propagators are obtained as series in $\hbar$, with
effective mass and coupling determined by non-perturbative gap
equations. This provides a systematic approximation method for
the ground state energy, and for the second and fourth
excitations. The results show quick convergence to the exact
values, better than that obtained without including the operator
$\Phi^4$.
\end{abstract}
\vskip 4 cm
\newpage

\section{Introduction}
The quantum mechanical anharmonic oscillator (AO) can be
regarded as a quantum field theory of a real scalar field with a
classical action given by
\begin{equation}
S[\Phi]=\int\![\frac{1}{2}\Phi(x)(-\partial^2+m^2)\Phi(x)+
\lambda\Phi^{4}(x)]\,d^{n}x,
\label{Scl}
\end{equation}
when the dimension of the Euclidean space-time $n$ is equal to
one. For this reason the AO is frequently used as a testing
ground for approximation methods in quantum field theory. The
field-theoretical perturbation method enables us to calculate
Green's functions and their generating functionals to an
arbitrary order in the coupling constant $\lambda$. Within this
approach the perturbative expansion of the ground state energy
of the AO is easily obtained by calculating the effective
potential in the loop expansion, but a derivation of excitation
energies from zero modes of the appropriate Green's functions is
not so straightforward~\cite{AOGf}, in contrast to the
Schr\"odinger approach, where perturbative calculations for
excited levels are of the same difficulty as for the ground
state.

For a field-theoretical investigation of excited modes, the
$N-$particle functionals which depend on some artificial sources
coupled to composite operators,
$\Phi(x_{1})\Phi(x_{2})...\Phi(x_{N})$, have been
introduced~\cite{LY,DM}. These functionals generate composite
propagators through differentiation with respect to the sources.
The simplest Green's functions, with the simplest relations to
multi-particle eigenmodes, are generated by the effective action
for composite operators, defined through a Legendre transform of
the connected generating functional. For $N\le 4$ the effective
action has been shown~\cite{DM,Vas} to be a sum of
$N$-particle-irreducible vacuum diagrams with the exact
propagator. In the case $N=2$, studied in the classic paper of
Cornwall, Jackiw and Tomboulis~\cite{CJT}, the Gaussian
effective action (and the Gaussian effective
potential~\cite{GEP}) is obtained as an approximation of the
two-loop result. It is, however, difficult to discuss a full
content of the two-loop result, and to proceed to higher
operators within such an approach because of the highly
non-trivial integral gap equations for vacuum expectation values
of composite fields.

The local $N-$particle operators $\Phi^N(x)$ have been also
considered~\cite{Fuk}; however, diagrammatic rules for the
effective action were lacking, since the Legendre transform
cannot be performed explicitly in this case. Only recently, an
expansion of the effective action for the operator $\Phi^2(x)$
in terms of two-particle-point-irreducible (2PPI) diagrams was
given~\cite{chin,ver}. The result is implicit but enables us to
calculate the effective potential~\cite{ver} and the
two-particle composite propagator~\cite{AOver}. The Gaussian
effective action is obtained already in the one-loop
approximation, and the calculation of post-Gaussian corrections
is easier than for the bilocal operator, since the gap equation
for the vacuum expectation value of the local composite field is
algebraic. Unfortunately this approach cannot be extended to
$N>2$, and all one can do in that case is to perform the
inversion in the Legendre transform order by order in the Planck
constant $\hbar$. For multilocal operators this uphill task can
be done explicitly, and the well-known expansions in
$N-$particle-irreducible diagramms are recovered~\cite{Oku}. For
the local composite operator $\Phi^2(x)$ the method appears
implicit; nevertheless it makes a discussion of many
applications possible~\cite{verinv,inv}. Even the diagrammatic
rules for the effective action (different from that of the 2PPI
expansion) have been established~\cite{Oku,Yoko}. The aim of our
paper is to extend the inversion method to include the composite
operator $\Phi^4(x)$.

In Section~\ref{inv}, after a brief exposition of the generating
functionals for composite operators, a diagrammatic
representation of the effective action for the $\Phi^2(x)$
operator is obtained explicitly to the order $\hbar^5$, and the
(inverse) two-particle composite propagator is derived up to
$\hbar^4$. Then we calculate the effective action for two
operators, $\Phi^2(x)$ and $\Phi^4(x)$, to the order $\hbar^4$.
This enables us to obtain the two- and four-particle propagators
up to $\hbar^3$ and $\hbar^2$, respectively. Having in mind an
application of the formalism in relativistic quantum field
theory, we discuss the scalar theory in the space-time of
$n$-dimensions, keeping $n$ arbitrary as long as possible. It is
only in Section~\ref{QM}, when calculating excitation energies,
that we set $n=1$ to consider the AO. The poles of the composite
propagators are determined in successive orders of $\hbar$ and
the resulting energies are compared with the exact spectrum of
the oscillator. The approximations for the ground state energy
and for the second excitation, obtained from the effective
action for $\Phi^2(x)$, are in good agreement with the exact
results; however, in this approach there is no simple way to
derive the fourth excitation. The advantage of using the
effective action for two operators, $\Phi^2(x)$ and $\Phi^4(x)$,
lies in the fact that it generates directly the four-particle
propagator, providing an excellent approximation to the fourth
excitation energy of the AO. Moreover, the approximations to the
ground state and to the second excitation obtained from the
effective action for the operators $\Phi^2(x)$ and $\Phi^4(x)$
are considerably better than that obtained from the effective
action for the operator $\Phi^2(x)$ only. Our conclusions are
summarized in Section~\ref{con}.

\section{The effective action for local composite operators
through the inversion method}
\label{inv}
\subsection{Generating functionals for composite operators} 
The vacuum functional for the local composite operators
$\Phi^2(x)$ and $\Phi^4(x)$ can be represented by a path
integral
\begin{equation}
Z[J,K]=e^{\frac{1}{\hbar}W[J,K]}=\int\!
D\Phi\,e^{\frac{1}{\hbar}\left[-S[\Phi] +
\frac{1}{2}\int\!\!J(x)\Phi^{2}(x)\,d^{n}x+
\frac{1}{24}\int\!\!K(x)\Phi^{4}(x)\,d^{n}x\right]}.
\label{Z4}
\end{equation}
The connected generating functional, $W[J,K]$, is given by a sum
of vacuum diagrams in configuration space, obtained with an
inverse propagator
$G^{-1}_{J}(x,y)=(-\partial^2+m^2-J(x))\delta(x-y)$ and a
four-point vertex given by
\mbox{$(K(x)-24\lambda)\delta(x-y)\delta(x-w)\delta(x-z)$}; in
Fig.~\ref{WW4} we show the diagrammatic expansion to the order
$\hbar^5$. For simplicity, we consider here the case of unbroken
reflection symmetry, when the vacuum expectation value of the
scalar field vanishes; this is why we do not introduce a current
coupled to an elementary field $\Phi(x)$. This, however, will
limit our investigation to the even excitations of the system.

The effective action is defined through a Legendre transform
\begin{eqnarray}
\Gamma[\Delta,\Lambda]&=&W[J,K]-\frac{\hbar}{2}\int\! J(x)
\Delta(x)\,d^{n}x -\frac{\hbar^3}{24}\int\! K(x)
\Lambda(x)\,d^{n}x\nonumber\\
&-&\frac{\hbar^2}{8}\int\! K(x)\Delta^2(x)\,d^{n}x,
\label{Gam}
\end{eqnarray}
where the background fields are defined as
\begin{eqnarray}
\hbar\Delta(x)&=&2 \frac{\delta W}{\delta J(x)}=<\Phi^2(x)>_{J,K}\nonumber\\
\hbar^3\Lambda(x)&=&24 \frac{\delta W}{\delta K(x)}-3\hbar^2\Delta^2(x)=
<\Phi^4(x)>_{J,K}-3<\Phi^2(x)>_{J,K}^2
\label{dJK}
\end{eqnarray}
and $<...>_{J,K}$ denotes the expectation value in the presence
of external currents $J(x)$ and $K(x)$. 

The external sources $J$ and $K$ were introduced artificially in
order to define the background fields~(\ref{dJK}) and the
effective action as a functional of those~(\ref{Gam}). The
physical quantities have to be calculated at $J=K=0$, or
equivalently at the values of the background fields, $\Delta$
and $\Lambda$, for which the gap equations
\begin{equation}
\frac{\delta\Gamma}{\delta \Delta(x)}=-\frac{\hbar}{2}J(x)-
\frac{\hbar^2}{4}\Delta(x) K(x)=0,
\label{J0}
\end{equation}
\begin{equation}
\frac{\delta\Gamma}{\delta \Lambda(x)}=-\frac{\hbar^3}{24}K(x)=0
\label{K0}
\end{equation}
are satisfied. The effective action for the composite operators,
$\Gamma[\Delta,\Lambda]$, determines the vacuum energy density
\begin{equation}
E_{0}=-\left.\frac{\Gamma[\Delta,\Lambda]}{\int\!
d^{n}x}\right|_{\Delta=const,\Lambda=const}.
\label{E0}
\end{equation}
If another solution to the gap equations ($\Delta(x)=\Delta+
D\Delta(x)$, $\Lambda(x)=\Lambda+ D\Lambda(x)$) exists in the
vicinity of ($\Delta$, $\Lambda$), then by~(\ref{J0})
and~(\ref{K0}) the stability condition~\cite{Fuk,Fuksta} is
obtained in the form
\begin{eqnarray}
\begin{array}{cc}
\int \left|\begin{array}{cc}
\Gamma_{22}(x,y)&\Gamma_{24}(x,y)\\
\Gamma_{42}(x,y)&\Gamma_{44}(x,y)\\
 \end{array}\right|
   \left|\begin{array}{c}
   D\Delta(y)\\D\Lambda(y),
   \end{array}\right|
\end{array}
dy=\left|\begin{array}{c}
   0\\0
   \end{array}\right|.
\label{g44}
\end{eqnarray}
The matrix $\Gamma$, given by
\begin{eqnarray}
\Gamma_{22}(x,y)&=&\left.\frac{\delta^2 \Gamma}{\delta\Delta(x)\delta\Delta(y)}
 \right|_{\Delta,\Lambda}=
-\left.\frac{\hbar}{2}\frac{\delta\left(J(x)+\frac{\hbar}{2}K(x)\Delta(x)
\right)}{\delta\Delta(y)}\right|_{\Delta, \Lambda}\nonumber\\
\Gamma_{24}(x,y)&=&\Gamma_{42}(x,y)=\left.\frac{\delta^2\Gamma}
{\delta\Delta(x)\delta\Lambda(y)}\right|_{\Delta,\Lambda}=
-\left.\frac{\hbar^3}{24}\frac{\delta K(x)}{\delta\Delta(y)}\right|_{\Delta, \Lambda}\nonumber\\
\Gamma_{44}(x,y)&=&\left.\frac{\delta^2 \Gamma}{\delta\Lambda(x)\delta\Lambda(y)}
 \right|_{\Delta,\Lambda}=
-\left.\frac{\hbar^3}{24}\frac{\delta K(x)}
{\delta\Lambda(y)}\right|_{\Delta, \Lambda},
\label{gg}
\end{eqnarray}
is an inverse of the composite propagator matrix, defined as
\begin{eqnarray}
W(x,y)&=&\left|\begin{array}{cc}
W_{22}(x,y)&W_{24}(x,y)\\
W_{42}(x,y)&W_{44}(x,y)\\
 \end{array}\right|=
\left|\begin{array}{cc}
\left.\frac{\delta^2 W}{\delta J(x)\delta J(y)}\right|_{0,0}&
\left.\frac{\delta^2 W}{\delta J(x)\delta K(y)}\right|_{0,0}\\
\left.\frac{\delta^2 W}{\delta K(x)\delta J(y)}\right|_{0,0}&
\left.\frac{\delta^2 W}{\delta K(x)\delta K(y)}\right|_{0,0}
\end{array}\right|\nonumber\\&=&\left|\begin{array}{cc}
<T\Phi^2(x)\Phi^2(y)>_{con}&<T\Phi^2(x)\Phi^4(y)>_{con}\\
<T\Phi^4(x)\Phi^2(y)>_{con}&<T\Phi^4(x)\Phi^4(y)>_{con}
\end{array}\right|.
\label{ww}
\end{eqnarray}
In Fourier space the stability condition~(\ref{g44}) is written as
\begin{eqnarray}
\begin{array}{cc}
\left|\begin{array}{cc}
\Gamma_{22}(p)&\Gamma_{24}(p)\\
\Gamma_{42}(p)&\Gamma_{44}(p)\\
 \end{array}\right|
   \left|\begin{array}{c}
   D\Delta(p)\\D\Lambda(p),
   \end{array}\right|
\end{array}
=\left|\begin{array}{c}
   0\\0
   \end{array}\right|,
\label{g44p}
\end{eqnarray}
hence the zero modes of $\Gamma(p)$ are related to the
excitations of the system. The matrix $\Gamma(p)$ can be
diagonalised, providing the inverse of the two-particle
propagator in the form
\begin{eqnarray}
\hbar\Gamma^{4}(p)=\Gamma_{22}(p)-\Gamma_{24}(p)\Gamma^{-1}_{44}(p)
\Gamma_{42}(p)
\label{gg2}
\end{eqnarray}
and that of the four-particle propagator equal to
\begin{eqnarray}
\hbar^3\Gamma^{8}(p)=\Gamma_{44}(p)-\Gamma_{42}(p)\Gamma^{-1}_{22}(p)
\Gamma_{24}(p).
\label{gg4}
\end{eqnarray}

The effective action for the composite operators,
$\Gamma[\Delta,\Lambda]$, provides a convenient tool to study
the ground state and excitations. Unfortunately, a systematic
approximation scheme for the effective action is intricate,
although the connected generating functional, W[J,K], can be
easily obtained in terms of Feynman diagrams in the loop
expansion. The main difficulty lies in performing the Legendre
transform~(\ref{Gam}), i.e. in eliminating the sources $J$ and
$K$ in favour of $\Delta$ and $\Lambda$. Here we shall do this
order by order in $\hbar$, representing the vacuum expectation
values $\Delta$ and $\Lambda$ by the series
\begin{equation} 
\Delta=\sum_{k=0}^{\infty}\hbar^k \Delta_{(k)}[J,K],
\label{dJp}
\end{equation}
\begin{equation} 
\Lambda=\sum_{k=0}^{\infty}\hbar^k \Lambda_{(k)}[J,K],
\label{dKp}
\end{equation}
where the coefficients $\Delta_{k}[J,K]$ and $\Lambda_{k}[J,K]$
can be easily found from Eq.~\ref{dJK}. The diagrammatic
representation of $\Delta$ and $\Lambda$ to the order $\hbar^4$
and $\hbar^2$ respectively is given in Fig.~\ref{DL}. The
inverted series are written as
\begin{equation} 
J=\sum_{k=0}^{\infty}\hbar^k J_{(k)}[\Delta,\Lambda],
\label{ddep}
\end{equation}
\begin{equation} 
K=\sum_{k=0}^{\infty}\hbar^k K_{(k)}[\Delta,\Lambda],
\label{dlap}
\end{equation}
and the coefficients $J_{(k)}[\Delta,\Lambda]$ and
$K_{(k)}[\Delta,\Lambda]$ have to be determined order by order
in $\hbar$. It has to be noted that only the lowest-order
inverse functionals, $J_{0}[\Delta,\Lambda]$ and
$K_{0}[\Delta,\Lambda]$, are required; higher order coefficients
are obtained as functionals of them. This establishes an
algorithm for calculating the effective action to an arbitrary
order in $\hbar$ even if $J_{0}[\Delta,\Lambda]$ and
$K_{0}[\Delta,\Lambda]$ cannot be obtained explicitly, which is
the case for local composite operators.

\subsection{The effective action for $\Phi^2(x)$, $\Gamma[\Delta]$}
\label{inv2}
The effective action for the two-particle operator $\Phi^2(x)$,
$\Gamma[\Delta]$, has been recently examined with great care by
Okumura~\cite{Oku} and Yokojima~\cite{Yoko}. As an introduction
to the case of four-particle operator, discussed in the next
section, we review here the calculation of $\Gamma[\Delta]$ by
performing the Legendre transform of the connected generating
functional $W[J]$ to the order $\hbar^5$. In the definition of
the effective action for the two-particle operator,
$\Gamma[\Delta]$, only the current $J(x)$ coupled to $\Phi^2(x)$
is required; therefore $K(x)$ has to be set to zero in the
formulas for the generating functionals~(\ref{Z4}-\ref{dde}) and
only Eq.~(\ref{dJp}) has to be inverted, to determine a
functional $J[\Delta]$ to the given order in $\hbar$. Using the
identity
\begin{equation} 
\Delta=\sum_{k=0}^{\infty}\hbar^k
\Delta_{(k)}[\sum_{i=0}^{\infty}\hbar^i J_{(i)}[\Delta]],
\label{ide}
\end{equation}
one obtains an infinite sequence of equations
\begin{equation} 
\Delta=\Delta_{(0)}
\label{ide0}
\end{equation}
\begin{equation} 
\Delta_{(0)}^{'}J_{(1)}+\Delta_{(1)}=0,
\label{ide1}
\end{equation}
\begin{equation} 
\Delta_{(0)}^{'}J_{(2)}+\frac{1}{2}\Delta_{(0)}^{''}(J_{(1)})^2+
\Delta_{(1)}^{'}J_{(1)}+\Delta_{(2)}=0,
\label{ide2}
\end{equation}
\begin{eqnarray} 
\lefteqn{\Delta_{(0)}^{'}J_{(3)}+\Delta_{(0)}^{''}J_{(1)} J_{(2)}}\nonumber\\
&+&\frac{1}{3!}\Delta_{(0)}^{'''}(J_{(1)})^3+\Delta_{(1)}^{'} J^{(2)}
+\frac{1}{2}\Delta_{(1)}^{''} (J_{(1)})^2+\Delta_{(3)}=0,
\label{ide3}
\end{eqnarray}
\begin{eqnarray} 
\lefteqn{\Delta_{(0)}^{'}J^{(4)}+\frac{1}{2}\Delta_{(0)}^{''}(2J_{(1)}J^{(3)}
+(J_{(2)})^2)}\nonumber\\&+&\frac{1}{2}\Delta_{(0)}^{'''}(J_{(1)})^2 J_{(2)}+
\frac{1}{4!}\Delta_{(0)}^{''''}(J_{(1)})^4+\Delta_{(1)}^{'}J^{(3)}
+\frac{1}{3!}\Delta_{(1)}^{'''}(J_{(1)})^3\nonumber\\
&+&\Delta_{(1)}^{''} J_{(1)} J_{(2)}+\Delta_{(2)}^{'} J_{(2)}+\frac{1}{2}
\Delta_{(2)}^{''}(J_{(1)})^2+\Delta_{(3)}^{'}J_{(1)}+\Delta_{(4)}=0~~~
\label{ide4}
\end{eqnarray}
\begin{displaymath}
...,
\end{displaymath}
where $\Delta_{(k)}[J_{0}]$ are given in Fig.~\ref{DL}. Here and
in the following the space-time indices and integration over
them are suppressed for notational simplicity. The coefficients
$J_{(i)}$, calculated from Eqs.~\ref{ide1}-~\ref{ide4}, can be
represented by Feynman diagrams in configuration space with the
(inverse) free propagator
\begin{equation}
G^{-1}_{J_{0}}(x,y)=(-\partial+\Omega^2(x))\delta(x-y),
\label{prop}
\end{equation}
where
\begin{equation}
\Omega^2[\Delta]=m^2-J_{0}[\Delta].
\label{om}
\end{equation}
The result to the order $\hbar^4$ is shown in Fig.~\ref{Ji}; the
inverse two-particle propagator, attached to an external point,
is that of two free particles of an effective mass $\Omega(x)$.
In Fig.~\ref{Gam2} we show the effective action $\Gamma[\Delta]$
to the order $\hbar^5$, obtained by eliminating $J$ with the use
of Fig.~\ref{Ji}. In agreement with the rule proved by
Okumura~\cite{Oku}, $\Gamma[\Delta]$ appears to be a sum of
one-vertex-irreducible diagrams with the inverse composite
propagator attached to an external point, and the two-point
pseudo-vertices $J_{(i)}$ of order $i$ ($i\ge 2$) inserted in
all possible ways. By way of digression, it can be observed that
all pseudo-vertex insertions in $\Gamma[\Delta]$ can be summed
by taking a dressed inverse propagator in the form
$G^{-1}_{sum}(x,y)=
(-\partial^2+m^2-J_{0}(x)-J_{2}(x)-J_{3}(x)-J_{4}(x)+...)\delta(x-y)=
(-\partial^2+m^2-J(x)+J_{1}(x))\delta(x-y)$, and the result
coincides with that obtained in the 2PPI expansion~\cite{ver}
which establishes a relation between the two methods.

The definition of the effective action $\Gamma[\Delta]$ is
implicit, since the functional
\begin{equation}
\Delta[J_{0}](x)=G_{J_{0}}(x,x),
\label{Gxx}
\end{equation}
obtained from~(\ref{ide0}), cannot be inverted for a space-time
dependent $J_{0}$. However, using the fact that all dependence
on $J$ is through $J_{0}$,~Eq.\ref{Gxx} enables us to calculate
$\left.\frac{\delta
J_{0}(z)}{\delta\Delta(y)}\right|_{\Delta(x)=\Delta}$, and hence the
two-particle inverse propagator
\begin{equation}
\Gamma^{4}(x-y)=\Gamma_{22}(x-y)=
-\left.\frac{\hbar}{2}\int d^{n}z \frac{\delta J(x)}{\delta
J_{0}(z)}\frac{\delta J_{0}(z)}{\delta
\Delta(y)}\right|_{\Delta(x)=\Delta}.
\label{g4}
\end{equation}
Since a constant $\Delta$ corresponds to constant $\Omega$, the
vacuum energy density~(\ref{E0}), as well as the Fourier
transform of the two-particle two-point Green's function,
$\Gamma^{4}(p)$, can be represented by the Feynman diagrams in
momentum space with the propagator $\frac{1}{p^2+\Omega^2}$. The
vacuum energy density is given by the same set of diagrams as
the effective action~(Fig.~\ref{Gam2}), but in momentum space.
The result for $\Gamma^{4}(p)$ calculated from~(\ref{g4}) to the
order $\hbar^3$ is shown in Fig.~\ref{pro2}. In this
case~(\ref{ide0}) can be written as
\begin{equation}
\Delta=\int \frac{d^{n}p}{(2\pi)^{n}}\frac{1}{p^2+\Omega^2},
\label{Dconst}
\end{equation}
and can be used to eliminate $\Delta$ in favor of $\Omega$ in
all the quantities considered. The gap equation~(\ref{J0}),
which becomes algebraic, determines the value of $\Omega$ which
corresponds to vanishing external current.

\subsection{The effective action for $\Phi^2(x)$ and $\Phi^4(x)$
operators, $\Gamma[\Delta,\Lambda]$}
\label{inv4}
In order to calculate the effective action for the operators
$\Phi^2(x)$ and $\Phi^4(x)$, $\Gamma[\Delta,\Lambda]$, both the
series given by~(\ref{dJp}) and~(\ref{dKp}) have to be
inverted. Using the identities
\begin{equation} 
\Delta=\sum_{k=0}^{\infty}\hbar^k
\Delta_{(k)}[\sum_{i=0}^{\infty}\hbar^i J_{(i)}[\Delta,\Lambda],
\sum_{k=0}^{\infty}\hbar^i K_{(i)}[\Delta,\Lambda]].
\label{jde}
\end{equation}
\noindent and
\begin{equation} 
\Lambda=\sum_{k=0}^{\infty}\hbar^k
\Lambda_{(k)}[\sum_{i=0}^{\infty}\hbar^i J_{(i)}[\Delta,\Lambda],
\sum_{i=0}^{\infty}\hbar^i K_{(i)}[\Delta,\Lambda]].
\label{kde}
\end{equation}
we obtain two infinite sequences of equations
\begin{equation} 
\Delta=\Delta_{(0)}
\label{jde0}
\end{equation}
\begin{equation} 
\frac{\delta\Delta_{(0)}}{\delta J_{0}} J_{1}+\Delta_{(1)}=0,
\label{jde1}
\end{equation}
\begin{equation} 
\frac{\delta\Delta_{(0)}}{\delta J_{0}}J_{(2)}+
\frac{1}{2}\frac{\delta^2\Delta_{(0)}}
{\delta J_{0}^2}(J_{(1)})^2+
\frac{\delta\Delta_{(1)}} {\delta J_{0}} J_{(1)}+
\frac{\delta\Delta_{(1)}} {\delta K_{0}} K_{(1)}+\Delta_{(2)}=0,
\label{jde2}
\end{equation}
\begin{displaymath}
...,
\end{displaymath}
\noindent and
\begin{equation} 
\Lambda=\Lambda_{(0)}
\label{kde0}
\end{equation}
\begin{equation} 
\frac{\delta\Lambda_{(0)}}{\delta J_{0}}J_{1}+
\frac{\delta\Lambda_{(0)}}{\delta K_{0}}K_{1}+\Lambda_{(1)}=0,
\label{kde41}
\end{equation}
\begin{equation} 
\frac{\delta\Lambda_{(0)}}{\delta J_{0}}J_{(2)}+
\frac{1}{2}\frac{\delta^2\Lambda_{(0)}}{\delta J_{0}^2}(J_{(1)})^2+
\frac{\delta\Lambda_{(0)}}{\delta K_{0}}K_{(2)}+
\frac{\delta^2\Lambda_{(0)}}{\delta J_{0}\delta K_{0}}J_{(1)}K_{(1)}+
\frac{\delta\Lambda_{(1)}} {\delta J_{0}} J_{(1)}+\Lambda_{(2)}=0,
\label{kde42}
\end{equation}
\begin{displaymath}
...,
\end{displaymath}
where we made use of the fact that $\Delta_{(0)}$ and
$\frac{\delta\Lambda_{(0)}}{\delta K_{0}}$ do not depend on
$K_{0}$. Taking $\Delta_{(k)}[J_{0},K_{0}]$ and
$\Lambda_{(k)}[J_{0},K_{0}]$ from Fig.~\ref{DL}, we calculated
the coefficients $J_{(i)}$ and $K_{(i)}$, from the above
equations. The results for $J$ and $K$ to the order $\hbar^3$
and $\hbar^2$ respectively are represented in Fig.~\ref{JiKi} by
Feynman diagrams in configuration space, where the composite
propagators are those of two or four free particles. The
effective mass in a free propagator,
$\Omega[\Delta]$~(\ref{om}), is the same as in the case of
$\Gamma[\Delta]$ considered in Section~\ref{inv2}, and the
four-point vertex is
$(K_{0}(x)-24\lambda)\delta(x-y)\delta(x-w)\delta(x-z)$. In
Fig.~\ref{Gam4} we show the effective action,
$\Gamma[\Delta,\Lambda]$ to the order $\hbar^4$, which is
obtained as a Legendre transform of the connected generating
functional by using Fig.~\ref{JiKi} to eliminate $J$ and $K$ in
favour of $\Delta$ and $\Lambda$. It seems that, as in the case
of $\Gamma[\Delta]$, graphical rules could be established, but
we do not disscuss this point here.

The expression for $\Gamma[\Delta,\Lambda]$ given in
Fig.~\ref{Gam4} is implicit, with the relations for
$J_{0}[\Delta,\Lambda]$ and $K_{0}[\Delta,\Lambda]$ given
by~(\ref{jde0}) and~(\ref{kde0}) respectively. The vacuum energy
density~(\ref{E0}) is represented by the same set of Feynman
diagrams as the effective action (Fig.~\ref{Gam4}), but in
momentum space with the propagator $\frac{1}{p^2+\Omega^2}$.
Using the fact that all dependence on sources is through $J_{0}$
and $K_{0}$, the implicit expression for
$\Gamma[\Delta,\Lambda]$ enables us to derive the inverse of the
composite propagators matrix, defined for constant $\Delta$ and
$\Lambda$, and its Fourier transform, $\Gamma(p)$ can be
diagonalized. In Figs.~\ref{pro42} and~\ref{pro44} we show the
Feynman diagram representations for the inverse of the
two-particle propagator, $\Gamma^4(p)$ (up to $\hbar^3$), and
for that of the four-particle propagator, $\Gamma^8(p)$ (up to
$\hbar^2$), respectively.

For constant $\Delta$ and $\Lambda$, $J_{0}$, $K_{0}$ and
$\Omega$ are also space-time independent, thus~(\ref{jde0})
becomes~(\ref{Dconst}), and~(\ref{kde0}) can be written as
\begin{eqnarray} 
\Lambda&=&\Lambda_{(0)}(J_{0},K_{0})\nonumber\\&=&(K_{0}-24\lambda)\int
\frac{d^{n}p d^{n}q
d^{n}r}{(2\pi)^{3n}(p^2+\Omega^2)(q^2+\Omega^2)(r^2+\Omega^2)
((p+q+r)^2+\Omega^2)}~~~~~
\label{Lconst}
\end{eqnarray}
and inverted to obtain
\begin{equation} 
K_{0}=24\lambda+\frac{\Lambda}{
\int
\frac{d^{n}pd^{n}qd^{n}r}{(2\pi)^{3n}(p^2+\Omega^2)(q^2+\Omega^2)
(r^2+\Omega^2)((p+q+r)^2+\Omega^2)}}.
\label{Kconst}
\end{equation}
The relations~(\ref{Dconst}) and~(\ref{Kconst}) can be used to
eliminate $\Delta$ and $K_{0}$ in favor of $\Omega$ and
$\Lambda$ in the calculated quantities. The values $\Omega$ and
$\Lambda$ have to be determined from the gap
equations~(\ref{J0}) and~(\ref{K0}), which become algebraic.

\section{The quantum-mechanical anharmonic \mbox{oscillator}}
\label{QM}
In the space-time of one dimension the $\lambda\Phi^4$ theory is
equivalent to the quantum mechanical anharmonic oscillator with
a Hamiltonian given by
\begin{equation}
H=\frac{1}{2} p^2+\frac{1}{2} m^2 x^2+\lambda x^4.
\label{AO}
\end{equation}
The exact element of the composite propagator matrix in the
Euclidean formulation can be represented as
\begin{eqnarray}
W_{i j}(t-t')&=&\ll\!0|Tx^{i}(t)x^{j}(t')|0\!\gg-\ll\!0|x^{i}(t)|0\!\gg
\ll\!0|x^{j}(t')|0\!\gg\nonumber\\&=&\sum_{k=1}^{\infty}
\ll\!0|x^i|k\!\gg \ll\!k|x^j|0\!\gg e^{-|t-t'|\varepsilon_{k}},
\end{eqnarray}
where $\epsilon_{k}=E_{k}-E_{0}$ is the excitation energy of the
$k-$th state, $|k\!\gg$. The Fourier transform of the above
propagator is given by
\begin{eqnarray}
W_{i j}(p)=\sum_{k=1}^{\infty}\frac{2\epsilon_{k}\ll\!0|x^i|k\!\gg
\ll\!k|x^j|0\!\gg } {p^2+\varepsilon_{k}^2},
\end{eqnarray}
and has an infinite number of poles at imaginary momenta whose
absolute values determine all excitations of the system.
Equivalently the full spectrum can be obtained by looking for
zero modes of the exact inverse propagator matrix, $\Gamma(p)$.
However, if we are working with an approximate propagator, which
may have only a finite number of poles, this provides an
approximation to some part of the energy spectrum only.
Different, but usually better, approximations can be obtained
from zero modes of the $\Gamma$ matrix, and such an approach
will be used in our work.

\subsection{The approximations of the AO spectrum}
\label{QM2}
\subsubsection{The approximations derived from $\Gamma[\Delta]$}
The vacuum energy density of the scalar field in the one
dimensional space-time gives the ground state energy of the AO.
Using the effective action for the operator $\Phi^2$,
$\Gamma[\Delta]$, given in Fig.~\ref{Gam2} the ground state energy
of the AO is calculated to be
\begin{equation}
E_{0} = \hbar\frac{\Omega}{2}-(\Omega^2 - m^2)\hbar\frac{\Delta}{2} 
+ 3\lambda\hbar^2\Delta^2 - \frac{3\lambda^2\hbar^3}{8\Omega^5} 
+\frac{27\lambda^3\hbar^4}{16\Omega^8}
-\frac{1923\lambda^4\hbar^5}{128\Omega^{11}}+ ...
\label{e20}
\end{equation}
where $\Delta$ is space-time independent, and by
Eq.~(\ref{Dconst} is given by
\begin{equation} 
\Delta=\frac{1}{2\Omega}.
\label{id}
\end{equation}
The two-particle inverse propagator, from Fig.~\ref{pro2},
becomes equal to
\begin{eqnarray}
-\Gamma^{4}(p)=\frac{\Omega}{2}(p^2+4\Omega^2)+6\lambda\hbar
-3\lambda^2\hbar^2\frac{160\Omega^2+p^2}{\Omega^3(16\Omega^2+p^2)}
+27\lambda^3\hbar^3\frac{7168\Omega^4+176\Omega^2p^2 + p^4}
        {2\Omega^6 (16\Omega^2+p^2)^2}\nonumber\\
-3\lambda^4\hbar^4\frac{5198561280\Omega^8+416746496\Omega^6 p^2+
      11879232\Omega^4 p^4+160692\Omega^2 p^6+641p^8}
      {16\Omega^9 (16\Omega^2+p^2)^3 (36\Omega^2+p^2)}+...~~~~
\label{pa2}
\end{eqnarray}
and its zero determined to the order $\hbar^4$ results in an
approximation to the second excitation given by
\begin{eqnarray}
\epsilon_{2}=2\Omega+\frac{3\hbar\lambda}{\Omega^2}-
\frac{87\hbar^2\lambda^2}{4\Omega^5}+\frac{2547\hbar^3\lambda^3}
{8\Omega^8}-\frac{401691\hbar^4\lambda^4}{64\Omega^{11}}+...
\label{e22}
\end{eqnarray}
The effective mass $\Omega$ has to be determined from the gap
equation~(\ref{J0}) which, after using~(\ref{id}), in the
one-dimensional space-time becomes
\begin{eqnarray}
32\Omega^{12}-32 m^2\Omega^{10}-192\lambda\hbar\Omega^9+
240\lambda^2\hbar^2\Omega^6-1728\lambda^3\hbar^3\Omega^3+
21153\lambda^4\hbar^4+...=0.
\label{qmg}
\end{eqnarray}
It is important to notice here that the non-perturbative
character of the approximations studied in this work is due to
non-perturbative treatment of the gap equation - working with an
effective action at a given order in $\hbar$ the gap equation
will be truncated to that order in $\hbar$ and solved after
setting $\hbar=1$. One can easily check that if one expanded the
solution of the gap equation $\Omega$ to the given order in
$\hbar$, the results for the ground state energy and for the
second excitation would coincide to that order in $\lambda$ with
energies obtained in perturbation theory for Schr\"odinger
equation. A field-theoretical derivation of the perturbative
result for the second excitation in the loop expansion of the
conventional effective action is considerably more laborious,
since more diagrams have to be evaluated~\cite{AOGf}.

In this approach approximations to higher excitations cannot be
obtained, because composite propagators for higher number of
particles cannot be derived directly from $\Gamma[\Delta]$.
There is however another strategy, analogous to Brillouin-Wigner
perturbation theory, which seems to provide a way out, by
calculating all zeros in the expression for $\Gamma^{4}(p)$ of
this order, and treating them as approximations to successive
excitations. We have checked that the lowest zero of
$\Gamma^{4}(p)$ provides a good approximation to the second
excitation, and in this case the result is close to~(\ref{e22}).
However, the other zeros do not agree with the exact results:
the higher the excitation, the worse the agreement. Even the
perturbative behavior of the excitations, obtained by using a
solution of the gap equation expanded to the given order in
$\hbar$, is wrong. In fact, the Brillouin-Wigner strategy
applied to the two-particle inverse propagator is not able to
provide reasonable approximations for higher excitations and
does not show any supremacy over the Rayleigh-Schr\"odinger
strategy. Therefore, in our work we use the
Rayleigh-Schr\"odinger strategy, approximating excitation
energies by zeros of the composite propagators calculated to the
considered order in $\hbar$. In this approach higher excitations
can be inverstigated by considering the effective actions for
higher composite operators which directly generate higher
composite propagators.

\subsubsection{The approximations derived from $\Gamma[\Delta,\Lambda]$}
We shall derive an approximation to the AO spectrum from the
effective action for $\Phi^2(x)$ and $\Phi^4(x)$,
$\Gamma[\Delta,\Lambda]$, discussed in Section~(\ref{inv4}).
Calculating the Feynman diagrams of Fig.~\ref{Gam4} in the
space-time of one dimension the ground state energy is obtained
in the form
\begin{equation}
E_{0} = \hbar\frac{\Omega}{2}-(\Omega^2 - m^2)\hbar\frac{\Delta}{2} 
+ 3\lambda\hbar^2\Delta^2 + \frac{2}{3}\Lambda^2\hbar^3\Omega^5 
+\hbar^3\lambda\Lambda - 4\Lambda^3\hbar^4\Omega^7+...,
\label{e40}
\end{equation}
where~(\ref{id}) can be used to eliminate $\Delta$ in favor of
$\Omega$. The inverse two-particle propagator, taken from
Fig.~\ref{pro42}, in one-dimensional space-time becomes
\begin{eqnarray}
\Gamma^{4}(p)&=&\frac{\Omega}{2}(p^2+4\Omega^2)+6\lambda\hbar
-16\Lambda^2\Omega^7\hbar^2\frac{160\Omega^2+p^2}{3(16\Omega^2+p^2)}
\nonumber\\&+& 64\hbar^3\Lambda^3\Omega^9\frac{256\Omega^4+
176\Omega^2p^2+p^4}{(16\Omega^2+p^2)^2}+...,
\label{pa42}
\end{eqnarray}
and its zero, calculated to order $\hbar^3$, determines the
second excitation to be
\begin{eqnarray}
\epsilon_{2}=2\Omega+\frac{3\lambda\hbar}{\Omega^2}
-104\frac{\Lambda^2\hbar^2\Omega^5}{3}-\frac{9\lambda^2\hbar^2}{4\Omega^5}
-96\Lambda^3\hbar^3\Omega^7+20\Lambda^2\hbar^3\Omega^2\lambda
+\frac{27\lambda^3\hbar^3}{8\Omega^8}+...
\label{e24}
\end{eqnarray}
The inverse four-particle propagator, from Fig.~\ref{pro44},
takes the form
\begin{eqnarray}
\Gamma^{8}(p)&=&2\Omega^3(16\Omega^2 +p^2)+\hbar(32\Omega^5-576\Lambda\Omega^7
 +2\Omega^3 p^2-12\Lambda\Omega^5 p^2)\nonumber\\&+&\frac{\hbar^2}
   {9(4\Omega^2 +p^2)(36\Omega^2 + p^2)}\left( 41472\Omega^9
 -746496\Lambda\Omega^{11}-331776\Lambda^2\Omega^{13}\right.\nonumber\\&
 +&14112\Omega^7 p^2 - 222912\Lambda\Omega^9 p^2 
 + 544384\Lambda^2\Omega^{11} p^2 
 +1008\Omega^5 p^4-9504\Lambda\Omega^7 p^4\nonumber\\&+&\left.  
32768\Lambda^2\Omega^9 p^4 +18\Omega^3 p^6 - 108\Lambda\Omega^5 p^6 
+ 88\Lambda^2\Omega^7 p^6\right)+...
\label{pa44}
\end{eqnarray}
and its zero, calculated to order $\hbar^2$, determines the
fourth excitation to be
\begin{eqnarray}
\epsilon_{4}=4\Omega- 24\Lambda\hbar\Omega^3
-\frac{560\Lambda^2\hbar^2\Omega^5}{3}+...
\label{e44}
\end{eqnarray}
In the above expressions the effective mass, $\Omega$, and the
effective coupling, $\Lambda$, are space-time independent and
have to be determined from the gap equations~(\ref{J0})
and~(\ref{K0}), which in one-dimensional space-time become
\begin{equation}
m^2-\Omega^2-8\hbar\Lambda\Omega^4+\frac{176\hbar^2\Lambda^2\Omega^6}{3}
-\frac{400\hbar^3\Lambda^3\Omega^8}{9}+...=0
\label{gapj}
\end{equation}
\begin{equation}
24\hbar\lambda+32\hbar\Lambda\Omega^5-288\hbar^2\Lambda^2\Omega^7+
\frac{5632\hbar^3\Lambda^3\Omega^9}{9}...=0
\label{gapk}
\end{equation}
When considering the approximations to the AO spectrum at a
given order, we shall truncate these equations to that order in
$\hbar$, and solve them after setting $\hbar=1$. If one expanded
instead the solutions of the gap equations, $\Omega$ and
$\Lambda$, in powers of $\hbar$, the perturbative results for
the ground state energy and the second excitation to order
$\lambda^3$ and for the fourth excitation to order $\lambda^2$
would be recovered.

\subsection{The numerical results for the AO energy levels}
Here we shall compare the numerical results for the AO spectrum,
obtained from $\Gamma[\Delta]$, with those from
$\Gamma[\Delta,\Lambda]$. With the use of $\Gamma[\Delta]$ only
the ground state and the second excitation can be calculated,
but $\Gamma[\Delta,\Lambda]$ enables us to obtain in addition the
fourth excitation. Both methods are non-perturbative, since the
gap equations truncated to the given order in $\hbar$ have been
solved numerically, after setting $\hbar=1$. In both cases the
solution with the largest positive value for $\Omega$ has been
chosen. The results are compared with the exact values of the AO
energy levels, calculated by a numerical procedure based on
a modification of the linear variational method~\cite{AOAO},
and with the results of perturbation theory. All results are
presented as functions of the dimensionless quantity
$z=\frac{m^2}{2\lambda^{2/3}}$, which is the only parameter of
the theory after rescaling all quantities in terms of
$\lambda$.

In Fig.10 we present the results for the ground state energy to
successive orders in $\hbar$. The quality of an approximation
can be expressed in terms of a critical value of the parameter
$z$, below which large discrepancies between results of
different orders and the exact value appear. The critical value
for approximations~(\ref{e20}) obtained from $\Gamma[\Delta]$,
$z_{crit}\approx-1.$, appears much smaller than that for the
perturbative results, $z_{crit}\approx2.$. The
approximations~(\ref{e40}), obtained from
$\Gamma[\Delta,\Lambda]$, are still better hence the critical
value is still smaller, $z_{crit}\approx-3.$

The results for the second excitation, shown in Fig.11, are very
similar to those for the ground state, only the critical values
are correspondingly larger: $z_{crit}\approx4.$ for the
perturbative calculations, $z_{crit}\approx2.$ for the
approximations obtained from $\Gamma[\Delta]$~(\ref{e22}), and
$z_{crit}\approx-.5$ for that obtained from
$\Gamma[\Delta,\Lambda]$~(\ref{e42}). These results should
be compared with those obtained in the same order of the 2PPI
expansion~\cite{AOver}. We observe that the two methods of
expanding $\Gamma[\Delta]$ provide the approximations to the AO
spectrum of a similar quality.

The approximations for the fourth excitation can be obtained
from the inverse four-particle propagator $\Gamma^8(p)$, which
can be derived directly from $\Gamma[\Delta,\Lambda]$, but not
from $\Gamma[\Delta]$. As can be seen in Fig.12 the
results~(\ref{e44}), obtained from $\Gamma[\Delta,\Lambda]$, are
in good agreement with the exact result for the values of the
parameter $z$ greater than $z_{crit}\!\approx\!1$. This should
be compared with the perturbative results where
$z_{crit}\!\approx\!5$.

It is interesting to observe that $\Gamma[\Delta]$ and
$\Gamma[\Delta,\Lambda]$ provide reasonable results for ground
state energy even in the case of the double-well anharmonic
potential ($z<0$), if the wells are not too deep. Since the
results derived from $\Gamma[\Delta,\Lambda]$ are considerably
better, one may suppose that including higher composite operator
could improve the convergence properties and make the method
applicable even in the case of deeper wells. In all methods
considered the numerical results for the excitations are worse
than that for the ground state: the higher the excitation, the
larger the critical values below which the method does not
converge, but in any case including the operator $\Phi^4$
improves the convergence. One can speculate that increasing the
number of composite operators included in the effective action,
we could make the method applicable for higher excitations in
the case of the double well potential.

\section{Conclusions}
\label{con}
In the quantum scalar field theory the effective action for the
local composite operator $\Phi^2(x)$, $\Gamma[\Delta]$, can be
obtained by performing the Legendre transform of the connected
generating functional $W[J]$ order by order in $\hbar$. This
provides a convenient tool to study the ground state and the
second excitation in the theory. The result to the order
$\hbar^2$ coincides with the Gaussian approximation, where the
effective propagator is a Hartree one; higher orders provide a
way to go beyond this result. A further improvement of the
method can be achieved by taking into consideration higher
composite operators $\Phi^{k}(x)$; their effects appear
successively in higher orders of $\hbar$. 

In this work we investigated the effects of including the
operator $\Phi^{4}(x)$ in the effective action,
$\Gamma[\Delta,\Lambda]$, which has been calculated as a series
in $\hbar$. The effects of the operator $\Phi^{4}(x)$ first
appear at order $\hbar^3$. The successive approximations for the
energy density and for the inverses of the two- and
four-particle propagators, $\Gamma^{4}(p)$ and $\Gamma^{8}(p)$
have been derived from $\Gamma[\Delta,\Lambda]$. To each order
in $\hbar$ the effective mass and coupling have to be determined
from the algebraic gap equations. The numerical results for the
ground state and lowest excitations have been calculated for the
theory in the space-time of one dimension, i.e., for quantum
mechanical anharmonic oscillator. In this case the gap equations
were solved numerically, and the excitation energies were
determined as zeros of the inverse composite propagators in
successive orders of $\hbar$. A comparison with the results
obtained from the effective action for the operator $\Phi^2(x)$,
$\Gamma[\Delta]$, shows that including the operator $\Phi^4(x)$
improves the convergence of the approximation to the ground and
the second excition considerably. Moreover, the effective
action $\Gamma[\Delta,\Lambda]$ enables us to obtain the
approximation to the fourth excitation which is not possible
with the use of $\Gamma[\Delta]$.

One may note that the Gaussian approximation, obtained as the
lowest approximation to the effective action for the operator
$\Phi^2(x)$, appears also in a variational calculation with the
Gaussian trial propagator. However, it is not staightforward to
obtain a substantial improvement beyond the result of this
simple ansatz by modification of the trial
propagator~\cite{Nev}. Including higher operators into the
effective action provides such a improvement, and in addition
enables one to study excitations of the system.

\newpage
\noindent {\bf{\Large Figure captions}}\\

\noindent Figure 1. The loop expansion of W[J,K] in terms of
Feynman diagrams in configuration space; the line denotes the
propagator $(-\partial^2+\Omega^2(x))^{-1}\delta(x-y)$, and the
full circle stands for the four-particle coupling
$(-24\lambda+K(x))\delta(x-y)\delta(x-w)\delta(x-z)$.\\

\noindent Figure 2. $\Delta[J,K]$ and $\Lambda[J,K]$ represented
with Feynman rules of Fig.1; the small empty circle denotes an
external point $x$.\\

\noindent Figure 3. $J[\Delta]$ obtained by inversion of
$\Delta[J]$ in powers of $\hbar$. The line denotes the
propagator $(-\partial^2+\Omega^2(x))^{-1}\delta(x-y)$ and the
four-particle vertex is
$-24\lambda\delta(x-y)\delta(x-w)\delta(x-z)$. Two lines meeting
at two points with the slash across represent the inverse of the
free two-particle propagator.\\

\noindent Figure 4. $\Gamma[\Delta]$ represented with Feynman
rules of Fig.3.\\

\noindent Figure 5. The inverse of the two-particle propagator,
calculated from $\Gamma[\Delta]$, represented in terms of Feynman
diagrams in momentum space; the line stands for a free
propagator $\frac{1}{p^2+\Omega ^2}$, the small empty circle denotes
an external momentum $p$.\\

\noindent Figure 6. $J[\Delta,\Lambda]$ and $K[\Delta,\Lambda]$
obtained by inversion of $\Delta[J,K]$ and $\Lambda[J,K]$ in
powers of $\hbar$; the single line denotes the propagator
$(-\partial^2+m^2-J_{0}(x))^{-1}\delta(x-y)$, the lines meeting
at two points represent the free composite propagator of the
corresponding number of fields, and the slash means the
inversion.\\

\noindent Figure 7. $\Gamma[\Delta,\Lambda]$ represented with
Feynman rules of Fig.6.\\

\noindent Figure 8. The inverse of the two-particle propagator,
calculated from $\Gamma[\Delta,\Lambda]$, represented in terms
of Feynman diagrams in momentum space; the line stands for a free
propagator $\frac{1}{p^2+\Omega^2}$, the small circle denotes an
external momentum $p$.\\

\noindent Figure 9. As Fig.8, but for the inverse of the
four-particle propagator, calculated from $\Gamma[\Delta,\Lambda]$.\\

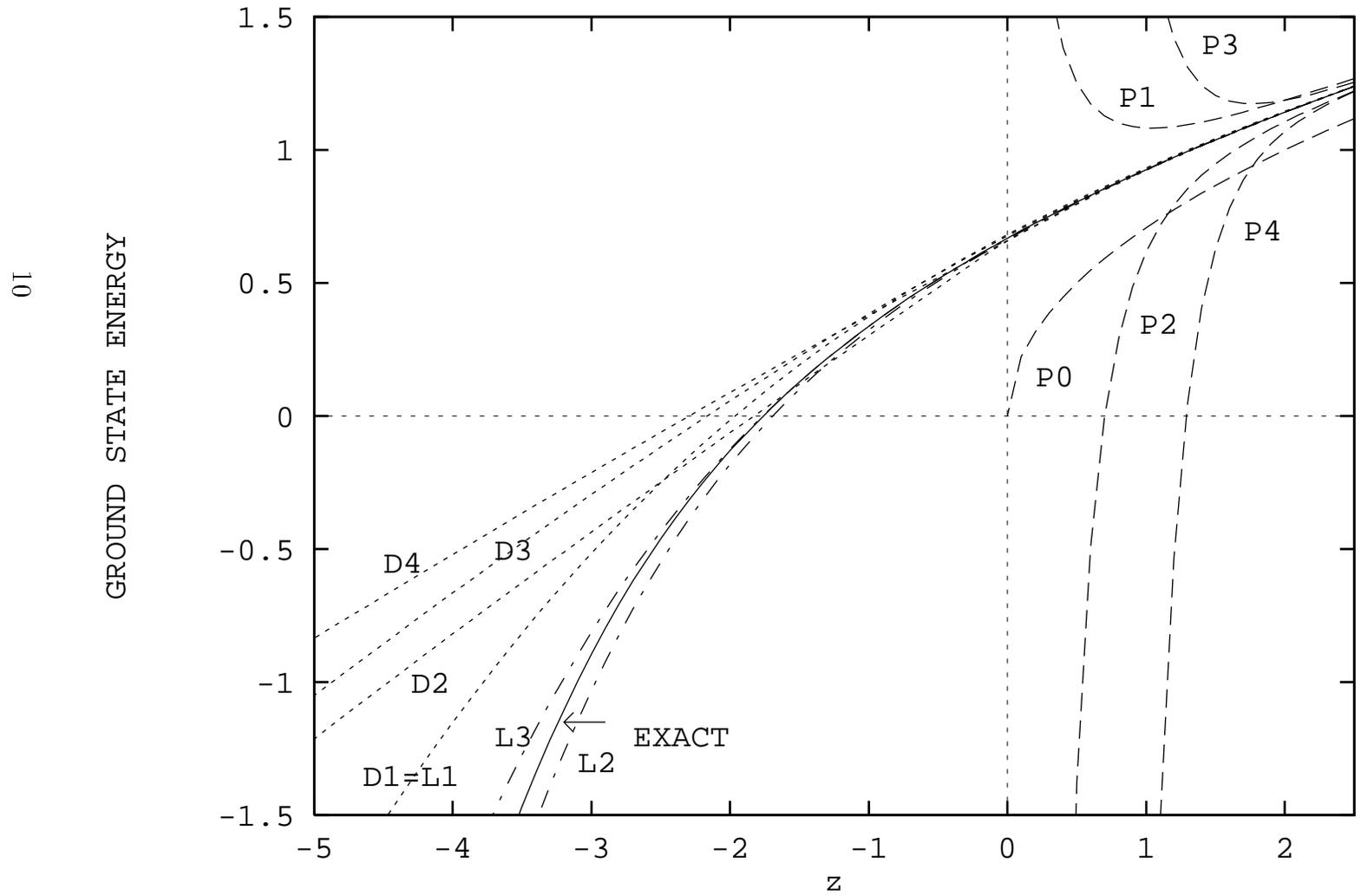
\noindent Figure 10. The ground state energy of the AO, obtained
to the given order of $\hbar$ from $\Gamma[\Delta]$ ({\it dotted
lines}) and from $\Gamma[\Delta,\Lambda]$ ({\it dotted-dashed
lines}), plotted {\it vs} $z=\frac{m^2}{2\lambda^{2/3}}$;
compared with the exact value ({\it solid lines}) and given
order perturbative results ({\it dashed lines}).\\

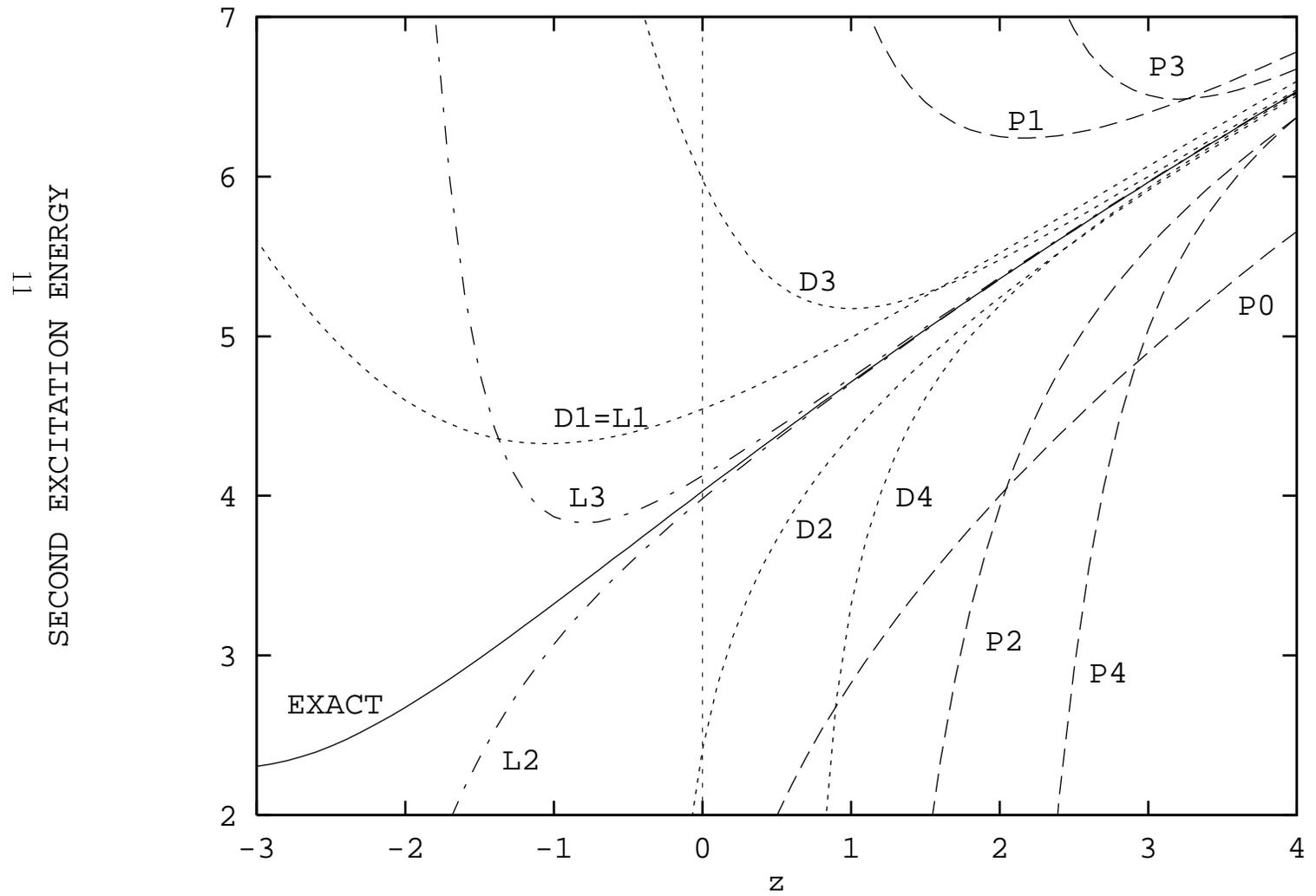
\noindent Figure 11. As in Fig.10, but for the second excitation
energy of the AO.\\

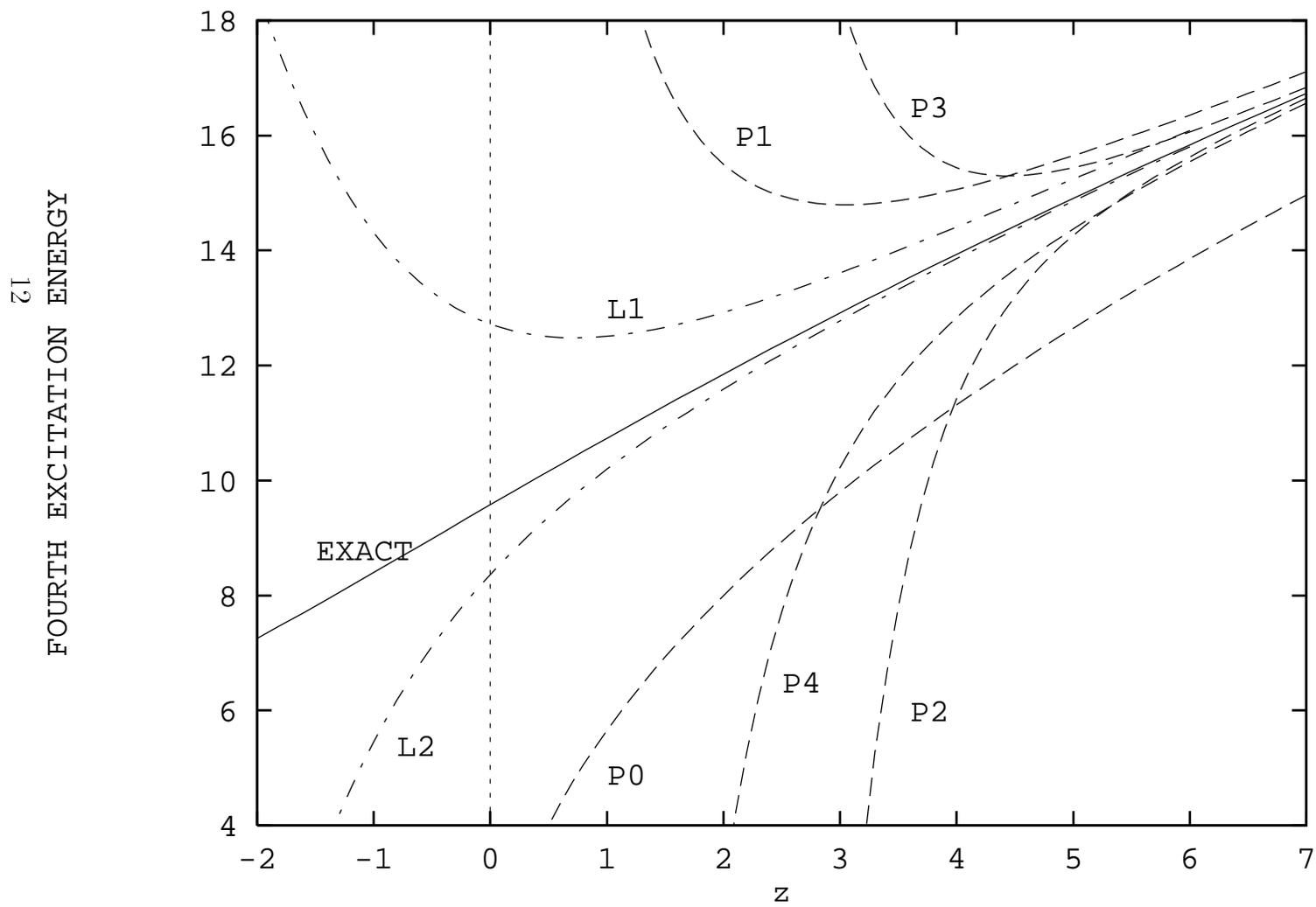
\noindent Figure 12. As in Fig.10, but for the fourth excitation
energy of the AO.\\
\end{document}